\documentclass[pra,reprint,showpacs,twocolumn,superscriptaddress,floatfix]{revtex4-1}

\usepackage{epsfig}
\usepackage{amsmath}
\usepackage{amssymb}
\usepackage{amsfonts}
\usepackage{mathptmx}
\usepackage{dcolumn}
\usepackage{eucal}
\usepackage{bm}
\usepackage{color}
\usepackage[colorlinks,linkcolor=blue,citecolor=blue]{hyperref}
\usepackage{epstopdf}
\usepackage{bbold}
\usepackage{url}
\usepackage{mathtools}

\usepackage{dsfont}

\newcommand{\Time}{\mathcal{T}}

\newcommand{\average}[1]{\langle #1 \rangle}

\def \ket#1{\mathinner{|{#1}\rangle}}

\newcommand{\ketbra}[2]{{\mathinner{| {#1} \rangle \langle {#2} |}} }

\newcommand{\matrixel}[3]{{\mathinner{\langle{#1}| {#2} | {#3}\rangle}} }

\newcommand{\GFuncF}{\mathcal{G}_{\chi,\mathcal{F}}}

\newcommand{\GFuncU}{\mathcal{G}_{\chi,\Delta U}}

\newcommand{\Prob}{\mathcal{P}}

\newcommand{\Idoperator}{\mathds{1}}

\def \tmp{TMP~}
\def \CharFun{QCGF~}
\def \ProbFun{QPDF~}

\begin{document}

\title{Quasi-probabilities of work and heat in an open quantum system}

\author{P. Solinas}
\affiliation{Dipartimento di Fisica, Universit\`a di Genova, via Dodecaneso 33, I-16146, Genova, Italy}
\affiliation{INFN - Sezione di Genova, via Dodecaneso 33, I-16146, Genova, Italy}
\author{M. Amico}
\thanks{Present address: Q-CTRL, Sydney, NSW Australia \& Los Ange- les, California $90013$, USA.
}
\affiliation{The Graduate School and University Center, The City University of New York, New York, NY 10016, USA}
\author{N. Zangh\`i}
\affiliation{Dipartimento di Fisica, Universit\`a di Genova, via Dodecaneso 33, I-16146, Genova, Italy}
\affiliation{INFN - Sezione di Genova, via Dodecaneso 33, I-16146, Genova, Italy}

\date{\today}

\begin{abstract}
We discuss an approach to determine averages of the work, dissipated heat and  variation of internal energy of an open quantum system driven by an external classical field.
These quantities are measured by coupling the quantum system to a quantum detector at different times.
This approach allows us to preserve the full quantum features of the evolution.
From the measured phase, we are able to obtain a quasi-characteristic function and a quasi-probability density function for the corresponding observables.
Despite the fact that these quasi-probability density functions are not the results of direct measurements, they reproduce the expected value of the physical quantities.
Analogously to the Wigner function, the negative regions of these quasi-probability density functions are directly related to pure quantum processes which are not interpretable in classical terms.
We use this feature to show that in the limit of strong dissipation, the quantum features vanish and interpret this as the emergence of the classical limit of the energy exchange process.
Our analysis explains and confirms the behavior observed in recent experiments performed on IBMQ devices \cite{solinas2021}.
The possibility to discriminate between classical and quantum features makes the proposed approach an excellent tool to determine if, and in which conditions, quantum effects can be exploited to increase the efficiency in an energy exchange process at the quantum level.
\end{abstract}

\maketitle

\section{Introduction}
The concepts of work and heat are keystones of classical physics.
They describe how a physical system exchanges energy with an external field and dissipates energy because of the interaction with an environment.

The recent development of quantum technologies has brought an unprecedented precision in controlling quantum systems allowing us to envision  quantum devices with novel capabilities to store and manipulate energy.
Therefore, it has become of utmost importance to understand how the classical concepts of work and heat are translated and changed in the quantum regimes.

This apparently simple research plan has found several roadblocks which arise from the fundamental aspects of quantum mechanics.
These can be exposed in the simplest possible situation of a closed quantum system subject to an external driving field.
In this case, the work done on the system is equal to the variation of the internal energy of the system.
However, we need information about both the initial and final energy. 
Since work and heat are related to a {\it process}  it is impossible to introduce a physically relevant hermitian operator as it would be non-local in time \cite{talkner2007fluctuation, Perarnau-Llobet2017No-Go, Lostaglio2015}.

A straightforward solution to avoid this problem is to measure the energy of the system at the beginning and at the end of the evolution.
This approach, often called Two Measurement Protocol (\tmp) \cite{campisi2011colloquium,campisi2011erratum}, has the advantage to be direct and to have a clear interpretation \cite{engel2007jarzynski} but it has several disadvantages as well.
First, it induces the wave-function collapse and, consequently,  changes the system energy.
Therefore, it could be tricky to separate the contribution to the work related to the external field (that we are interested in) from the one due to the (unwanted) perturbation induced by the detector.
Second, it changes the dynamics of the quantum system and, in general, modifies the expected average work done on the system because of the destruction of the interference contributions to the dynamics \cite{feynman1965quantum, solinas2013work,solinas2015fulldistribution,solinas2016probing}.
Thus, if we want to answer the question, "how much energy we need to run a quantum device",  the \tmp is unable to give us the correct answer.

Being forced to abandon the procedure generally used to define the measurement of a physical observable, i.e., introducing a corresponding hermitian operator, the clearest and direct approach to the problem is to include the detector in the dynamics, specify how it interacts with the system and how we obtain the desired information. 
This allows us to clearly identify what are its effects on the dynamics, which are the physical observables measured in the laboratory and how we can extract the desired information from the experimental measurements.

With this in mind, in this paper, we focus  on the approach suggested in Refs. \cite{ solinas2013work,solinas2015fulldistribution,solinas2016probing}. 
The key idea is that by coupling the system and the detector with a specific protocol, the information about the work, heat and internal energy can be stored in the phase of a quantum detector.
This was recently implemented on the IBMQ devices \cite{IBM_docs} to experimentally measure for the first time the average work, heat and internal energy  in a driven quantum system and to observe the reaching of the classical regimes of the quantum energy exchanges  \cite{solinas2021}.

Despite the correct interpretation of the experimental data in Ref. \cite{solinas2021}, a full theoretical framework is still missing since the analysis in presence of decohenrece phenomena has not been developed yet.
To fill this gap, we extend the results of Refs. \cite{ solinas2013work,solinas2015fulldistribution,solinas2016probing} to the case of open quantum dynamics by introducing the interaction with the environment.

From the measurement of the phase of the detector, we can calculate a {\it quasi-characteristic generating functions function} (\CharFun) and {\it quasi-probability density function} (\ProbFun) of the work, heat and internal energy.
As the prefix "quasi" points out, these probabilities are not obtained by direct experimental measurement and  indeed, the \ProbFun is, in general, not positively defined \cite{solinas2016probing, SolinasPRA2017}.
Still, as we show, in this way we can obtain the answer to our initial question determining the expected average work and heat that are naturally expected with no perturbation.

Interestingly, the \CharFun and \ProbFun carry important and deep information about energy exchanges.
In a direct analogy to the Wigner function \cite{WignerPhysRev1932}, the negative regions of the derived \ProbFun are associated with the violation of the Leggett-Garg inequalities and are the signature of a pure quantum phenomenon \cite{solinas2016probing}.

To complement and clarify this point, we calculate of the full \ProbFun of work, heat and internal energy in presence of an environment modeled as a quantum channel.
We are able to show how in the limit of strong dissipation the negative regions of the QPDFs disappear.
Interpreting this as the vanishing of pure quantum features of the process, we conclude that the environment tends to make the evolution classical.
In other words, the protocol can also be used to observe the emergence of the classical limit in the energy exchanges of a quantum system.

The discussed approach is particularly interesting for the experimental implementation and determination of the dissipated heat.
In fact, the alternative theoretical proposal \cite{campisi2011colloquium} was based on the extension of the \tmp and requires the direct projective measurement of the environmental degrees of freedom.
Since the environment generally has a large number of degrees of freedom and we do not have access to them, this proposal is not useful in practice.

An alternative proposal suggests to measure the quanta emitted by the system \cite{pekola2013calorimetric}.
Although closer to realistic implementation \cite{pekola2013calorimetric, Gasparinetti2015, Viisanen_2015, Saira2016, Vesterinen2017, Karimi2020, Karimi2020NatCom}, these methods are indissolubly tied to a specific physical implementation.
On the contrary, being based on the system degrees of freedom, the proposed approach can be implemented in any physical quantum system.
This practical advantage could have important experimental implications.

The paper has the following structure.
In Sec. \ref{section:work_definition} we recall what is a reasonable way to define the average work done on a closed system by an external time-dependent drive.
In Sec.  \ref{sec:QPDF_approach} we summarise the main ideas and formalism of Refs. \cite{solinas2015fulldistribution,solinas2016probing,SolinasPRA2017}.
We develop this approach to calculate the QPDFs for an open system in Sec. \ref{sec:QPDF_open_system} and discuss the main properties in Sec. \ref{sec:QPDF_properties}.
Section \ref{sec:conclusions} contains the conclusions.

\section{Natural definition of work}
\label{section:work_definition}

Let us consider a quantum system subject to a classical time-dependent drive. The Hamiltonian is denoted by $H_S(t)$ and suppose that the desired evolution occurs for $0 \leq t \leq \Time$.
The Hamiltonian eigenstates and eigenvalues are $\ket{n_t}$  and $\epsilon_{n_t}$, respectively.
Let us focus on the closed system case with no dissipation and suppose that the system is initially described by the density operator $\rho^0$.
The average initial internal energy is $\average{\epsilon_i} = \sum_{n_0} \rho_{n_0 n_0}^0 \epsilon_{n_0}$ where $\rho_{n_0 n_0}^0= \matrixel{n_0}{\rho_0}{n_0}$ and $\ket{n_0}$ is the eigenstate with energy $\epsilon_{n_0}$.

The final density operator is $\rho^\Time = U \rho^0 U^\dagger$ and the final average energy is $\average{\epsilon_\Time} = {\rm Tr} \left[ H_\Time \rho^\Time \right] $: 
\begin{equation}
	\average{\epsilon_\Time}= \sum_{m_\Time,n_0} |U_{m_\Time n_0}|^2 \rho_{n_0 n_0}^0 \epsilon_{m_\Time} + \sum_{m_\Time, n_0\neq k_0} U_{m_\Time n_0} \rho_{n_0 k_0}^0 U^\dagger_{k_0 m_\Time} \epsilon_{m_\Time}
\end{equation}
with obvious notations for the matrix elements.
The work done on the system is naturally defined as the difference of the system energies, i.e., $\average{W}=\average{\epsilon_\Time}  - \average{\epsilon_0}$ \cite{solinas2013work, solinas2015fulldistribution},
\begin{eqnarray}
\average{W} &=& \sum_{m_\Time,n_0} |U_{m_\Time n_0}|^2 \rho_{n_0 n_0}^0 (\epsilon_{m_\Time} - \epsilon_{n_0})  \nonumber \\
&&+ \sum_{m_\Time, n_0\neq k_0} U_{m_\Time n_0} \rho_{n_0 k_0}^0 U^\dagger_{k_0 m_\Time} \epsilon_{m_\Time}.
 \label{eq:ideal_average}
\end{eqnarray}
Notice that this is the work done by the drive {\it with no influence or perturbation of any measurement apparatus} and, in this sense, it is the work done only by the drive on the system.

To point out the destructive effect of the measurement, we compare this result with the one predicted in the \tmp \cite{campisi2011colloquium,campisi2011erratum} in which all the moments of the work can be obtained with simple probabilistic arguments \cite{engel2007jarzynski}.
The average work reads
\begin{equation}
 \average{W} = \sum_{n_0} \rho_{n_0 n_0}^0 \sum_m |U_{mn}|^2 (\epsilon_{m_\Time} - \epsilon_{n_0})
 \label{eq:TMP_first_moment}
\end{equation}
where $ |U_{m_\Time n_0}|^2$ is the probability to have the transition $\ket{n_0} \rightarrow \ket{m_0}$ because of the dynamics induced by the operator $U$.

Equations (\ref{eq:ideal_average}) and (\ref{eq:TMP_first_moment}) differ for the contribution of the initial off-diagonal elements of the density matrix $\rho_{n_0 k_0}^0$.
These are not present in Eq. (\ref{eq:TMP_first_moment}) because are destroyed in the \tmp by the initial measurement of the energy.
However, if we expect the work with no perturbation to be the one in Eq. (\ref{eq:ideal_average}), these contributions are fundamental and give rise, in Feynmann's words \cite{feynman1965quantum}, to the quantum interference effects.

It can be shown that the alternative approach which exploits a quantum detector presented in Refs \cite{solinas2015fulldistribution,solinas2016probing} correctly predicts the average in Eq. (\ref{eq:ideal_average}) and, in this sense, it preserves all the information about the dynamics and the initial coherence without the perturbation of the measurement.

As a side note, we want to stress that the results in Eq. (\ref{eq:ideal_average}), i.e., the one predicted with the approach discussed below, coincides with the one we would obtain with the weak-value measurements approach \cite{weak_value1988}.
In the latter, the weak coupling between the system and the detector allows us to measure the average value of the desired observable with no disturbance on the system dynamics.

\section{Physical system and detector}
\label{sec:QPDF_approach}

To have a self-consistent discussion, we briefly recall the proposal to obtain the \ProbFun  of work, heat and internal energy \cite{solinas2015fulldistribution,solinas2016probing}.

We consider the quantum system described by the Hamiltonian $H_S(t)$ coupled with a quantum detector.
We denote with $H_D = \sum_{\lambda} \lambda \ketbra{\lambda}{\lambda}$ and $H_{SD} = f(\chi, t) H_S(t) \otimes H_D$ the detector and system-detector coupling Hamiltonian, respectively, and $\ket{\lambda}$ are eigenstate of the detector Hamiltonian.
The function $f(\chi, t)$ determines the times and the strength $\chi$ of the system-detector coupling.

The main idea of the approach discussed in Refs. \cite{solinas2015fulldistribution,solinas2016probing} is to couple the system and the detector at certain times to store the information about the energy in the detector phase.

To clarify this point, we first discuss the way to determine the variation of internal energy $\Delta U$.
Since $\Delta U$ is a state function, we need to know only the initial and the final energy.
This is done by coupling the system and the detector only at the beginning ($t=0$) and at the end of the evolution ($t=\Time$) and let the system evolves "freely" (with no detector interaction but under the time-dependent drive) for $0 < t< \Time$.
Formally, this corresponds choosing $f(\chi, t) =\chi [ \delta(\Time -t) - \delta(t)]$ which generates the system-detector evolution  \cite{solinas2015fulldistribution,solinas2016probing} 
\begin{equation}
 \mathcal{U}_{\chi, \Delta U} =U_{\chi,\Time} ~U_S~ U_{-\chi,0}
 \label{eq:deltaU_temp}
\end{equation}
where $U_{-\chi,0} = \exp{[ -i \chi H_S(0) \otimes H_D]}$  and $U_{\chi,\Time} = \exp{[i \chi H_S(\Time) \otimes H_D]}$.
The phase accumulated in the detector can be measured and represents the \CharFun of $\Delta U$ denoted by $\GFuncU$.
The derivatives of $\GFuncU$ with respect to the coupling $\chi$ are the quasi-moments of $\Delta U$, i.e., $\average{\Delta U^n} = (-i)^n d^n \GFuncU/d \chi^n |_{\chi=0}$ \cite{clerk2011full, bednorz2012nonclassical}.
In Refs \cite{solinas2015fulldistribution, solinas2016probing} it was shown that for a closed system this procedure can be used to obtain the average value expected from Eq. (\ref{eq:ideal_average}).
 
To measure the statistics of the dissipated heat and work we use the same ideas with a slightly more complex implementation.
In fact, the heat and work are not state functions and depend on the path swept by the system during the evolution.
This implies that we must repeatedly measure the system during the evolution to keep track of the dissipated heat.

With simple arguments, it can be shown that the work done on the quantum system is associated with the variation of $H_S(t)$ in time, while the dissipated heat is associated with the change of the density matrix due to dissipative dynamics \cite{solinas2013work}.
Since work and heat are associated to different physical processes, we exploit this fact by separating in time the two processes in order to be able to distinguish them.

We discretize the evolution in $N+1$ steps such that $t_s = s \Delta t $ and  $\Delta t \ll \Time$ (with $s$ integer and $0\leq s \leq N$).
Since the system Hamiltonian changes over time $\Time$, under this condition, in every time interval the system Hamiltonian can be considered constant.
From the energetic point of view, no work is done on the system and only the dissipative dynamics takes place.
Therefore, all the variation in energy of the system in this time interval can be interpreted as dissipated heat due to the interaction with the environment.
Assuming that the system-detector coupling time is much smaller than $\Delta t$, the information about this energy change can be stored in the detector with the same coupling scheme of Eq. (\ref{eq:deltaU_temp}).

More specifically, for any $\Delta t$, the total Hamiltonian at time $t_s$ is $H^s = H_S^s + H_{SE} + H_E$ with $U_{s} = e^{-i \Delta t H^{s} }$. 
For small enough $\Delta t$, we can write  $U_{s} \approx  e^{-i \Delta t H_{SE}  } e^{-i \Delta t H_E  } e^{-i \Delta t H^{s}_S }$.
In the Born approximation, we assume that the environment is large enough, i.e., with so many degrees of freedom, that it is not affected by the interaction with the system. This corresponds to assuming that the environment has no internal dynamics and the $\exp\{-i \Delta t H_E \}$ term in  $U_{s} $ has no effect.

Within each small time interval $\Delta t$, the Hamiltonian $H_S^s$ can be considered constant. At the beginning and at the end of each interval, we instantaneously couple the system and the detector, i.e., on time-scale over which the system does not evolve.
In analogy with (\ref{eq:deltaU_temp}), and since $[H^{s}_S,  H_S^s \otimes H_D]=0$, the evolution operator for each interval reads 
\begin{equation}
  \mathcal{U}_{\chi/2} ^{s} = e^{- i  \frac{\chi}{2} H_S^s \otimes H_D} e^{-i \Delta t H_{SE}  } e^{ i  \frac{\chi}{2} H_S^s \otimes H_D}  e^{-i \Delta t H^{s}_S }\ .
  \label{eq:heat_meas_block}
\end{equation}
Notice that unitary dynamics and the dissipative dynamics are factorized and with this scheme we are able to measure the energy variation of the system due to the dissipation, i.e., the heat.

Each $\mathcal{U}_\chi ^{s}$ is defined so that we keep track of the heat $Q_s$ dissipated in the time interval $(s-1) \Delta t \leq t \leq s \Delta t $. Because of the form of the system-detector coupling Hamiltonain, the information on the dissipated heat along the evolution is stored in the phase accumulated between eigenstates $\ket{\lambda}$ of the detector Hamiltonian.

Notice the sign in the exponents that takes into account the fact that an emission (absorption) by the environment, i.e., decreasing (increasing) of the environment energy, corresponds to an absorption (emission) process of the system, i.e., increasing (decreasing) of the system energy (see appendix \ref{app:heat_sign}).

In order to account for the work done, we must use another scheme and add another system-detector coupling at the beginning and end of the evolution \cite{solinas2015fulldistribution}.
Putting things together, the total evolution operator reads
\begin{equation}
 U_{\chi/2} = e^{i  \frac{\chi}{2} H_S^N \otimes H_D} \Pi_{s=0}^{N} \mathcal{U}_{\chi/2} ^{s} e^{-i  \frac{\chi}{2} H_S^0 \otimes \Sigma_z}. 
 \label{eq:full_U}
\end{equation}
In the case of unitary evolution, $H^s = H_S^s$ and we immediately recover the closed-system result for the variation of the internal energy.

\section{Quasi-characteristic generating functions and quasi-probabilities distributions in open quantum system}
\label{sec:QPDF_open_system}

As discussed above, the physical observable in the present approach is the detector phase.
To obtain this, we must calculate the system-environment dynamics, trace out the system degrees of freedom and extract the phase accumulated in the detector states.
This is associated with the quasi-characteristic generating function $\GFuncF =\matrixel{\lambda }{\rho_D(t) }{-\lambda }/\matrixel{\lambda }{\rho^0_D }{-\lambda }$ and its Fourier transform gives us the quasi-probability distribution function $\Prob(\mathcal{F}) = \int d \chi \GFuncF e^{i \chi \mathcal{F}}$ \cite{solinas2015fulldistribution, solinas2016probing}.

We use the prefix "quasi" to stress that both $\GFuncF$ and $\Prob(\mathcal{F})$ are not obtained by an experimental measurement but are obtained by the measured phases. 
Their interest lies in the fact that they allow us to obtain the desired average values of the observables (\ref{eq:ideal_average}) and, at the same time, preserve the full quantum features of the process.

As we will see below, this could be used as a tool to distinguish the classical from the quantum nature of an energy exchange process and to reveal the emergence of the classical limit when the system is coupled to an environment.

\subsection{Heat  quasi-probability density distribution}
\label{sec:Heat_PDF}

To describe the effect of the environment we use the operator sum-representation \cite{nielsen-chuang_book, Garcia-Perez2020}.
This is an effective phenomenological way to describe the dissipative dynamics induced by the $\exp\{-i \Delta t H_{SE}  \}$ operator that greatly simplifies the calculation.

For any time $t_s$ we define $k$ operator elements $\{M^{k,s}\}$ such that the dissipative evolution of the system density operator at any time interval $\Delta t$ is given by $\rho_S(t_{s+1}) = \sum_k M^{k,s} \rho_S(t_s) (M^{k,s})^\dagger$.

We suppose that the system and detector are initially described by the factorized density operator $\rho_0 = \rho_{S} \otimes \rho_{D}= \sum_{i_0, j_0, \lambda,\lambda'} \rho_{i_0 j_0}  \rho_{\lambda\lambda'} \ketbra{i_0, \lambda}{j_0,\lambda'}$ where $\rho_{i_0 j_0}$ and $\rho_{\lambda\lambda'}$ are the system and detector density matrix elements, respectively.
Since we want to study energy exchanges, the eigenbasis of $H_S^{s}$ is the privileged basis to decompose the evolution. We denote the time-dependent eigenstates as with $\{ \ket{i_s}\}$, $\{ \ket{j_s}\}$ , $\{ \ket{n_s}\}$ and $\{ \ket{m_s}\}$.
The full evolution generated by Eq. (\ref{eq:heat_meas_block}) reads (see Appendix \ref{sec:Heat_PDF})
\begin{eqnarray}
\rho_0  & \rightarrow &
\sum \rho_{\lambda\lambda'} e^{ i  \frac{\chi}{2} [\lambda (\epsilon_{i_0} - \epsilon_{n_0})- \lambda' (\epsilon_{j_0}- \epsilon_{m_0}) ]} \times\nonumber \\
&&M^{k,0}_{n_0i_0}  U^0_{i_0 i_0} \rho_{i_0 j_0} (U^0)^\dagger_{j_0 j_0}(M^{k,0})^\dagger_{j_0m_0} \ketbra{n_0,\lambda}{m_0,\lambda'} (U^0)^\dagger_{j_0 j_0}. \nonumber  \\
\label{eq:dissipative_first_step}
\end{eqnarray}
The contribution $M^{k,0}_{n_0i_0}  U^0_{i_0 i_0} = \mathcal{A}^{P_1}_{n_0 l} $ gives the probability amplitude for the system to undergo the transition path $P_1:i_0 \rightarrow n_0$. This is associated with the dissipated energy $ \epsilon_{n_0}- \epsilon_{i_0}$ taken into account in the exponential factor.
Analogously, the path $P_2: j_0 \rightarrow m_0 $ is transversed with probability amplitude $(U^0)^\dagger_{j_0 j_0}(M^{k,0})^\dagger_{j_0m_0}= (\mathcal{A}^{P_2}_{r  m_0})^\dagger$ and it is associated with the dissipated energy $ \epsilon_{m_0}- \epsilon_{j_0}$.

Generalizing this result to the dynamics discretized in $N$ steps and denoting with $n_{s-1} = i_0$ and $m_{s-1} = j_0$, we have (Appendix \ref{sec:Heat_PDF})
\begin{eqnarray}
\rho_N &=& \sum_{P_1,P_2} \rho_{\lambda\lambda'}  
e^{i  \frac{\chi}{2} [\lambda q_{P_1} - \lambda' q_{P_2} ]} \times  \nonumber \\
&& \mathcal{A}^{P_1}_{n_N i_0} \rho_{i_0 j_0} (\mathcal{A}^{P_2}_{j_0 m_N})^\dagger 
 \ketbra{n_N,p}{m_N,\lambda'}
\label{eq:rhoN_heat}
\end{eqnarray}
where we have collected the probability amplitude of path $P_1: i_0 \rightarrow n_0 \rightarrow i_1 \rightarrow .... \rightarrow n_N$ and path $P_2: j_0 \rightarrow m_0 \rightarrow j_1 \rightarrow .... \rightarrow m_N$ and the corresponding dissipated heat $q_{P_1} =\sum_{s }(\epsilon_{i_s} - \epsilon_{n_s})$ and $q_{P_2} = \sum_s (\epsilon_{j_s}- \epsilon_{m_s})$, respectively.

Following Refs. \cite{solinas2015fulldistribution,solinas2016probing}, to extract the accumulated phase in the detector, we trace out the system degrees of freedom and take $\GFuncF =\matrixel{\lambda }{\rho_D(t) }{-\lambda }/\matrixel{\lambda }{\rho^0_D }{-\lambda }$.
We obtain 
\begin{equation}
 \GFuncF = \sum_{P_1,P_2}  
e^{i  \frac{\chi}{2} \lambda ( q_{P_1} + q_{P_2} )} \mathcal{A}^{P_1}_{n_N i_0} \rho_{i_0 j_0}
(\mathcal{A}^{P_2}_{j_0 n_N})^\dagger.
\end{equation}
The quasi-probability density function of the dissipated heat is obtained by taking the Fourier transform
\begin{equation}
 \Prob(Q) = \sum_{P_1,P_2}  
 \mathcal{A}^{P_1}_{n_N i_0} \rho_{i_0 j_0}
(\mathcal{A}^{P_2}_{j_0 n_N})^\dagger \delta \Big[Q -\frac{\lambda}{2} (q_{P_1}+ q_{P_2} )\Big]
\label{eq:PDF_Q}
\end{equation}
where in the paths $P_1$ and $P_2$ the last point is fixed to $j_N=i_N$ because of the trace over the system degrees of freedom.

\subsection{Internal energy  quasi-probability density distribution}
\label{sec:U_PDF}

The same approach can be used to calculate the internal energy density probability distribution where the system-detector coupling is turned on only at the beginning and at the end of the evolution. The calculation is similar to the one for the heat but with no intermediate coupling and no intermediate phase factor accumulated.
Therefore, only the initial and the final phase factors remain in Eq. (\ref{eq:rhoN_heat}), i.e., 
$\lambda \epsilon_{i_0} - \lambda' \epsilon_{j_0}$ and $\lambda \epsilon_{i_N} - \lambda' \epsilon_{j_N}$, respectively.
The final density operator reads
\begin{eqnarray}
\rho_N &=& \sum_{P_1,P_2} \rho_{\lambda\lambda'}  
e^{ i  \frac{\chi}{2} [\lambda (\epsilon_{i_0} - \epsilon_{n_N})- \lambda' \sum_l (\epsilon_{j_0}- \epsilon_{m_N}) ]} \nonumber \\
&& \mathcal{A}^{P_1}_{n_N i_0} \rho_{i_0 j_0}
(\mathcal{A}^{P_2}_{j_0 m_N})^\dagger \ketbra{n_N,\lambda}{m_N,\lambda'}
\end{eqnarray}
Passing to the \CharFun and then to the QPDF, we obtain
\begin{equation}
 \Prob(\Delta U) = \sum_{P_1,P_2}  
 \mathcal{A}^{P_1}_{n_N i_0} \rho_{i_0 j_0}
(\mathcal{A}^{P_2}_{j_0 n_N})^\dagger \delta \Big[\Delta U -\lambda \left(\epsilon_{n_N} - \frac{\epsilon_{i_0}+ \epsilon_{j_0}}{2} \right)\Big]
\label{eq:PDF_U}
\end{equation}
This results generalizes the one in Refs. \cite{solinas2016probing} for an open system evolution.

If we consider a two-level system with an energy gap $\epsilon$, i.e., $\epsilon_{s}=\pm \epsilon/2$, for $\epsilon_{i_0}=-\epsilon_{j_0}$ and  $\epsilon_{n_N} =\pm \epsilon/2$, in Eq. (\ref{eq:PDF_U}) we have the exchanges of half energy quantum.
These are related to the interference effect between the two paths $P_1$ and $P_2$.
They are present only if the system is initially in a superposition of eigenstates of the energy so that the initial energy value is not well-defined.
It can be shown  that these are associated to a negative quasi-probability region and to the violation of Leggett-Garg inequalities \cite{solinas2015fulldistribution,solinas2016probing, hofer2016,Potts2019,Levy2020}.
Therefore, we consider the half quantum energy exchanges and the corresponding negative probability regions as pure quantum features.

\subsection{Work quasi-probability density distribution}
\label{sec:W_PDF}

The last step we need is the derivation of the work quasi-probability density distribution that can be obtained from the heat distribution with a small change in the protocol \cite{solinas2015fulldistribution,solinas2016probing}.

The same dynamical evolutions in Eqs. (\ref{eq:heat_meas_block}) and (\ref{eq:full_U}) can be rewritten highlighting the sequence of Hamiltonian changes alternated to the dissipative evolution
\begin{equation}
	U_{tot} = ... e^{ i  \frac{\chi}{2} H_S^2 \otimes H_D} e^{- i  \frac{\chi}{2} H_S^1 \otimes H_D} U_1 e^{ i  \frac{\chi}{2} H_S^1 \otimes H_D} e^{ i  \frac{\chi}{2} H_S^0 \otimes H_D} U_0 
\end{equation}
where the operator $U_k = \exp\{-i \Delta t H^s \}$ includes the full open dynamics with constant a Hamiltonian $H_S^s$.

The coupling sequence $\exp\{ i  \frac{\chi}{2} H_S^1 \otimes H_D\} \exp\{ i  \frac{\chi}{2} H_S^0 \otimes H_D \}$ stores in the detector the information about the work due to the change $H_S^0 \rightarrow H_S^1$.
However, if we look at the product $\exp\{- i  \frac{\chi}{2} H_S^1 \otimes H_D\}U_1 \exp\{ i  \frac{\chi}{2} H_S^1 \otimes H_D\}$, we recognize that the sequence that stores the information about the heat, i.e., the variation of energy due to the dissipative dynamics when the Hamiltonian is constant.
We conclude that the same system-detector schemes allow us to obtain information about the work done.

Alternatively, we can start from Eq. (\ref{eq:rhoN_heat}) and rewrite the sum in the exponent isolating the energies at the beginning as

\begin{eqnarray}
	\lambda q_{P_1} - \lambda' q_{P_2} &=&  \sum_{s } \Big[ \lambda (\epsilon_{i_s} - \epsilon_{n_s})- \lambda' (\epsilon_{j_s}- \epsilon_{m_s}) \Big] \nonumber \\
	&=& 
   \sum'_{s } \Big[ \lambda (\epsilon_{i_{s+1}} - \epsilon_{n_s} )- \lambda' (\epsilon_{j_{s+1}}- \epsilon_{m_s})  \Big] \nonumber \\
	&&   -(\lambda \epsilon_{i_0}-\lambda' \epsilon_{j_0})
\end{eqnarray}

The energy differences $\epsilon_{i_{s+1}} - \epsilon_{n_s}$ and $\epsilon_{j_{s+1}}- \epsilon_{m_s}$ are exactly the work done when the Hamiltonian changes from $H_S^s$ to $H_S^{s+1}$.
The last term $\lambda \epsilon_{i_0}-\lambda' \epsilon_{j_0}$ is obtained from the first coupling $\exp\{ i  \frac{\chi}{2} H_S^0 \otimes H_D\}$ and must be subtracted.
This means that the first detector-system coupling must be cancelled if we want the work statistics.
Analogously, we find that also the final coupling $\exp\{ i  \frac{\chi}{2} H_S^N \otimes H_D\}$ must be avoided \cite{solinas2015fulldistribution,solinas2016probing}.

From these examples, we arrive to the conclusion that the only difference between the protocols to measure the heat and the work done is in the first and the last system-detector coupling.
With these observations, we can write the quasi-probability density function for the work as 
\begin{widetext}
\begin{eqnarray}
 \Prob(W) &=& \sum_{P_1,P_2}  
 \mathcal{A}^{P_1}_{n_N i_0} \rho_{i_0 j_0} (\mathcal{A}^{P_2}_{j_0 n_N})^\dagger  \delta \Big[W -\frac{\lambda}{2} \left( \lambda \sum_{s }(\epsilon_{i_{s+1}} - \epsilon_{n_s} )- \lambda' \sum'_l (\epsilon_{j_{s+1}}- \epsilon_{m_s}) \right)\Big] \nonumber \\
&=& \sum_{P_1,P_2}  \mathcal{A}^{P_1}_{n_N i_0} \rho_{i_0 j_0}
(\mathcal{A}^{P_2}_{j_0 n_N})^\dagger 
\delta \Big[W -\frac{\lambda}{2} \left(q_{P_1}+ q_{P_2} - 2 \epsilon_{n_N} + \epsilon_{i_0}+ \epsilon_{j_0} \right)\Big]
\label{eq:PDF_W}
\end{eqnarray}
\end{widetext}
where in the last equation we have used the direct relation between the three distributions.
Indeed, notice that the contribution $2 \epsilon_{n_N} - \epsilon_{i_0}- \epsilon_{j_0}$ comes from neglecting the first and the last system-detector coupling and from tracing out the system degrees of freedom.

\section{Properties of the  quasi-probability density distributions} 
\label{sec:QPDF_properties}

\subsection{Averages and energy conservation}

We recall that the present method is built to reproduce the correct average value of the energetic observables.
Despite the possibility of building a quasi-probability density distributions, the interpretation of higher moments can be difficult. Therefore, we focus exclusively on the average values which have a clear and precise meaning.

If we write the generic form of the probability density distributions as 
\begin{equation}
 \Prob(\mathcal{F}) = \sum_{P_1,P_2} \mathcal{A}^{P_1}_{n_N i_0} \rho_{i_0 j_0} (\mathcal{A}^{P_2}_{j_0 n_N})^\dagger \delta(\mathcal{F} -f_{P_1,P_2})
\end{equation}
where $f_{P_1,P_2}$ are the discrete possible values that the variable $\mathcal{F}$ can take associated to the paths $P_1$ and $P_2$, its average value
\begin{equation}
  \average{\mathcal{F}} =  \int d \mathcal{F}~\mathcal{F}~ \Prob(\mathcal{F}) = \sum_{P_1,P_2} \mathcal{A}^{P_1}_{n_N i_0} \rho_{i_0 j_0} (\mathcal{A}^{P_2}_{j_0 n_N})^\dagger (f_{P_1,P_2}).
\end{equation}
 
Factorizing the probability amplitude contributions, by direct calculation we obtain (with $\lambda=1$)
\begin{equation}
 \average{\Delta U} + \average{Q}-\average{W}  = 0.
\end{equation}
This is the condition for the conservation of energy.
Notice that the energy is conserved not only on average but {\it along every path} of the evolution, i.e., every term in the summation vanishes.

\subsection{Limit of closed evolution}
\label{sec:closed_evolution}

It is interesting to calculate the above quasi-probability distribution functions in two important limiting cases of closed evolution and strong dissipation.

To simplify the discussion we consider a two-level system that undergoes a relaxation process from the excited to the ground state.
This is the prototypical dissipative process if we are interested in energy exchanges.
As above, we take $\epsilon_{s}=\pm \epsilon/2$.

We make two further assumptions.
First, the environment induces relaxation but cannot excite the system.
This occurs when the temperature of the environment is smaller than the system energy gap (divided by the Boltzmann constant).
Second, the dissipative processes are described as an amplitude-damping channel \cite{nielsen-chuang_book, Garcia-Perez2020}.
These assumptions allow us to explicitly write the dissipative operators $M^k$ (see Appendix \ref{app:strong_dissipation_dynamics}).

If there is no dissipation, we have that the dissipative operators $M^{k,s} = \Idoperator$; for example, $M^{k,0}_{i_0 n_0} = \delta_{i_0 n_0}$.
In Eq. (\ref{eq:dissipative_first_step}), we have that $i_0 = n_0$ and $j_0 = m_0$, the two system-detector couplings cancels out and no phase is accumulated.
This holds for any step of the evolution and implies that $q_{P_1} = q_{P_2} =0$.
Thus, the heat distribution  reads
\begin{equation}
 \Prob(Q) = \sum_{P_1,P_2}  
 \mathcal{A}^{P_1}_{n_N i_0} \rho_{i_0 j_0}
(\mathcal{A}^{P_2}_{j_0 n_N})^\dagger \delta (Q)
\end{equation}
and, as expected, it is centered around zero meaning that there is no dissipated heat.
At the same time, since $q_{P_1} = q_{P_2} =0$, from Eq. (\ref{eq:PDF_U}) and (\ref{eq:PDF_W}) we have that $\Prob(\Delta U) = \Prob(W)$ as it should be.
As discussed above, $\Prob(\Delta U)$ includes the exchanges of half quanta of energy that are related to the quantum feature of the process.
Therefore, in absence of environmental dissipation, these quantum features are preserved.

We can make a comparison between the predicted average work and the expected one in Eq. (\ref{eq:ideal_average}).
In the closed evolution limit, the probability amplitudes become a sequence of unitary evolution; we have (see Appendix \ref{sec:Heat_PDF}) $\mathcal{A}^{P_1}_{n_N i_0} = \Big( \Pi_{s=0}^N U^l_{i_s n_{s-1}}\Big) = U_S (\Time)$.
Separating the diagonal and off-diagonal contributions of the density operator, we have \cite{solinas2016probing,SolinasPRA2017} 
\begin{eqnarray}
&& \Prob(W) = \sum_{n_N,i_0}  
 |U_{n_N i_0}|^2 \rho_{i_0 i_0}\delta \Big[W -\lambda \left(\epsilon_{n_N} - \epsilon_{i_0}\right)\Big] \nonumber \\
&& + \sum_{n_N,i_0\neq j_0} 
 U_{n_N i_0} \rho_{i_0 j_0} U_{j_0 n_N}^\dagger \delta \Big[W -\lambda \left(\epsilon_{n_N} - \frac{\epsilon_{i_0}+ \epsilon_{j_0}}{2} \right)\Big].
\label{eq:PDF_U_unitary}
\end{eqnarray}
Calculating the average work $\average{W} = \int dW W \Prob(W)$ and taking into account that in the second sum $\epsilon_{i_0}+ \epsilon_{j_0} = 0$ because of the constraint $i_0\neq j_0$, we obtain Eq. (\ref{eq:ideal_average}).

A direct comparison with the prediction of \tmp in Eq. (\ref{eq:TMP_first_moment}) highlights that the difference lies in the quantum contributions in the second sum in Eq. (\ref{eq:PDF_U_unitary}) which is related to the quantum interference terms.

In this sense, the QPDFs contain more information. 
The obtained averages coincide when the system is measured initially or start from an energy eigenstate but differ in case the system is initially in a superposition of energy eigenstates.
In this latter case, the \tmp fails to predict the correct work done [Eq. (\ref{eq:ideal_average})] while the discussed approach allows us to preserve the information about the full quantum feature of the exchanged energy.

\subsection{Limit of strong dissipation. Emergence of the classical limit.}
\label{sec:strong_dissipation}

On the other side of the system-environment coupling strength, the limit of strong dissipation is particularly interesting.
This physically corresponds to the case in which the relaxation times are much smaller than the evolution times so that the system quickly relaxes to the {\it instantaneous} ground state of the Hamiltonian.

We consider the prototype case of a two-level system.
This restriction on the dimension of the system allows us to fully solve the dissipative dynamics since the dissipative operators $M^{k,s}$ are well known \cite{nielsen-chuang_book, Garcia-Perez2020} (see appendix \ref{app:strong_dissipation_dynamics}).
Despite this simplification, the main features of the energy exchange processes and the quasi-probaiblity distribution functions are preserved.
In addition, this case has been recently experimentally implemented in Ref. \cite{solinas2021}.

We denote the time-dependent ground and excited states of $H^s$ with $\ket{g_s}$ and $\ket{e_s}$, respectively, where $s$ is a time index and the corresponding energies with $\epsilon_{g_s}$ and $\epsilon_{e_s}$. 

In the strong dissipation limit, i.e., $p=1$, the effect of the interaction of the environment is to make the evolution classical. 
This corresponds in our case to the destruction of the interference effects that are typically associated to the quantum features \cite{solinas2015fulldistribution,solinas2016probing, solinas2021}.

This classical limit is conveniently analyzed by studying the behavior of $\Prob(\Delta U)$. 
It can be shown that (see Appendix \ref{app:strong_dissipation_dynamics}) in this case, it reads (assuming $\epsilon_{g_N} = \epsilon_{g_0}$)
\begin{equation}
  \Prob(\Delta U) = \rho_{g_0 g_0} \delta(\Delta U) +  \rho_{e_0 e_0} \delta \left[ \Delta U - (\epsilon_{e_0}-\epsilon_{g_N}) \right].
\end{equation}
Notice that since $\rho_{g_0 g_0}>0$ and $\rho_{e_0 e_0}>0$, this is a probability distribution function since it is a positive-definite function.

In addition, it can be interpreted in terms of classical trajectories and it would be the same that one would obtain by the TMP.
With probability $\rho_{g_0 g_0}$ the system starts and ends in the ground state. This process is associated with  no variation of internal energy. i.e., $\Delta U =0$.
With probability $\rho_{e_0 e_0}$ the system starts from the excited state and it ends in the ground state. This process is associated with  the energy variation is $\epsilon_{e_0}-\epsilon_{g_N}$.

This interpretation in terms of {\it classical} trajectories is only possible if there are no coherences \cite{solinas2015fulldistribution,solinas2016probing}. 
The half-energy exchanges as well as the negative probability regions that are present in the general expression (\ref{eq:PDF_U}) vanish in presence of strong dissipation.
Since all these features are associated to a quantum process and the violation of the Leggett-Garg inequality, we deduce the presence of the environment has made the evolution classical.
In other words, the absence of negative probability regions and half-energy exchanges are the signature of the emerging of the classical limit due to the presence of the environment. 

\subsection{The QPDFs contain the TMP distributions}
\label{sec:TMP_in_QPDF}

The use of a quantum detector allows to preserve the information about the quantum evolution and in particular the interference effects occurring when the system starts from a coherent superposition of energy eigenstates.
These eventually give the additional contributions the average as in Eq. (\ref{eq:ideal_average}) which are not present in the TMP averages [Eq. (\ref{eq:TMP_first_moment})]

Analogously, the QPDFs contain all the information about the dynamics of the system and the initial state that is lost in the TMP due to the initial measurement.
It is worth to discuss this point to highlight the possible use of the QPDF approach.

Let's consider first the limit of no dissipation discussed in Sec. \ref{sec:closed_evolution}.
The TMP distribution reads \cite{talkner2007fluctuation, Engel2007}
\begin{equation}
 \Prob_{TMP}(W) = \sum_{n_N,i_0}  
 |U_{n_N i_0}|^2 \rho_{i_0 i_0}\delta \Big[W -\lambda \left(\epsilon_{n_N} - \epsilon_{i_0}\right)\Big].
\end{equation}
This is exactly what we obtain from the QPDF (\ref{eq:PDF_U_unitary}) if we set to zero the off-diagonal density matrix elements $\rho_{lr}$.
Therefore, the present approach of building the QPDF is able to predict the correct distribution we would obtain with the TMP: it is sufficient to discard the contributions due to the $\rho_{lr}$ terms or, equivalently, the fractional quantum energy exchanges of $|\Delta \epsilon| = \epsilon/2$.

An analogous discussion can be done for the case of an open system.
Starting, for example, Eq. (\ref{eq:PDF_U}), we can separate the contribution related to the diagonal $\rho_{ll}$ and off-diagonal $\rho_{lr}$ density matrix elements.
Still, the two distributions coincide if we consider only the diagonal contributions.
This confirms that even in presence of dissipation, the QPDF contains all the information that can be extracted by the TMP.
Thus, in this sense, the QPDF is more general than the TMP.

\section{Conclusions}
\label{sec:conclusions}

We have discussed an alternative approach to determine the averages of work, heat and internal energy variation in a quantum system driven by an external field and interacting with an environment.
The use of a quantum detector allows us to preserve the quantum feature of the evolution and to obtain the expected averages of the physical observables.
This information can be retrieved from the phase accumulated by the detector.
Three different system-detector coupling schemes give us the average of the work done, the dissipated heat and the internal energy variation of the system.

Furthermore, we have built the quasi-probability distribution functions.
They contain much more information than the more common direct measurement, e.g., the two-measurement protocol.
As in the Wigner quasi-probability function \cite{WignerPhysRev1932}, this additional information is manifest in the presence of negative regions that are the signature of a quantum process.
In fact, these are directly related to the violation of the Leggett-Garg inequalities \cite{solinas2015fulldistribution, solinas2016probing, hofer2016, Potts2019, Levy2020}.

We have calculated the full QPDFs for an open system interacting with a dissipative amplitude damping channel and shown that they allow us to track all the energy exchanges in the process.
We have shown that the QPDFs correctly describe the physical process and the extracted physical quantities satisfy energy conservation.

Interestingly, we have shown that in the limit of strong dissipation, the quantum features vanish leading to the expected probability distribution obtained by direct measurement.
This result can be interpreted as the emergence of the classical limit of the process induced by the external environment.

The approach with the QPDFs has several advantages with respect to the TMP.
First, it reproduces the expected averages value of the physical observables [see Eq. (\ref{eq:ideal_average})].
Second, it preserves more information than the TMP. In fact, the latter distributions can be obtained by the QPDFs simply discarding the off-diagonal density matrix element contributions  [see, for example, Eq. (\ref{eq:PDF_U}), Eq. (\ref{eq:PDF_W}) and Sec.  \ref{sec:TMP_in_QPDF}].
Third, it has a great practical and experimental advantage (as shown recently \cite{solinas2021}) since only the system degrees of freedom are involved in the process.
For the measurement of the heat two main approaches have been pursued.
The first one imagines the direct projective measurement of {\it all} the degrees of freedom of the environment to reproduce the environment density matrix \cite{campisi2011colloquium,campisi2011erratum,talkner2007fluctuation,talkner2016}.
Since this is an impossible task, this approach is nothing but a theoretical interest.
The second suggests to measure the energy quanta emitted by the system \cite{pekola2013calorimetric}.
The main limitation in this case is that it works only for very specific systems \cite{ Gasparinetti2015, Viisanen_2015, Saira2016, Vesterinen2017, Karimi2020, Karimi2020NatCom} where only a channel is involved in the dissipative process.
On the contrary, the present approach allows us to determine the dissipated heat by only manipulating the system-detector coupling leading to a protocol that does not depend on the physical platform used as a quantum system.
This is clearly illustrated by a recent experimental implantation in an all-purpose quantum computer  \cite{solinas2021}.

All  these features make the discussed approach a privileged tool to study the energy exchange process at quantum level.
As the Wigner quasi-probability function has allowed us to arrive at a deeper understanding of quantum phenomena, the QPDFs could allow us to identify the situations in which the quantum effects play a key role in the energy exchange.
This would be the first step toward a more complete understanding of the energy exchange processes at the quantum level.
In fact, it is not yet clear if the exploiting of pure quantum effects can lead to an advantage in the energy exchange.
This is a prerequisite to envision and build energy-efficient quantum devices that could be extremely important for future technologies.

\begin{acknowledgments}
PS and NZ acknowledge financial support from INFN.
The views expressed are those of the authors and do not reflect the official policy or position of Q-CTRL.
\end{acknowledgments}

\appendix
\setcounter{equation}{0}

\section{Notation of the heat sign}
\label{app:heat_sign}

The coupling sequence $U_{\chi, t_i} ~U~ U_{-\chi, t_j}$ where $U_{\pm \chi,t} = \exp\{ i \chi H_S(t) \otimes H_D \}$, allows us to determine the variation of the internal energy of the system, i.e., $\Delta U = \epsilon_f - \epsilon_i$ \cite{solinas2015fulldistribution,solinas2016probing}.
This is the energy {\it supplied to the system}; if the system absorbs energy $\Delta U>0$ and if emits energy, $\Delta U<0$.
This can be seen directly by calculating the CGF and the average value as discussed in Ref. \cite{solinas2015fulldistribution}.

On the contrary, the transformation  $U_{-\chi,t_i} ~U_{diss}~ U_{\chi, t_j}$
measures the energy  {\it supplied by the system} to the environment.
In other words we are "measuring"  $q = \epsilon_i - \epsilon_f$.
Therefore, if $q=\epsilon_i - \epsilon_f>0$, the system emits an energy quantum decreasing its energy while the environment increases its energy-absorbing an energy quantum.
If $q=\epsilon_i - \epsilon_f<0$, the system absorbs an energy quantum increasing its energy while the environment decreases its energy emitting an energy quantum.

Notice that this is the opposite notation with respect to the usual one where the $q$ is the energy {\it supplied to the system} as heat.
As a consequence, the energy conservation law takes the form $ \average{\Delta U} + \average{Q}-\average{W} =0$.

\section{Heat  probability density distribution}
\label{sec:Heat_PDF}

We use the operator sum-representation \cite{nielsen-chuang_book, Garcia-Perez2020}.
This is an effective phenomenological way to describe the dissipative dynamics induced by the $\exp\{-i \Delta t H_{SE}  \}$ operator. By using the operator sum-representation, we can simplify the calculation effectively describing the evolution of the open system neglecting the complex environment dynamics.

The $k$ operator elements describing the dissipation process are denoted with $\{M^{k,s}\}$ where the $s$ is a time index. The density operator evolution from time $t_s$ to $t_{s+1}$ is given by $\rho_S(t_{s+1}) = \sum_k M^{k,s} \rho_S(t) (M^{k,s})^\dagger$.
Since the problem is time dependent, we need a different set of operator elements $\{M^{k, l}\}$ where $0 \leq l \leq N$ is now a new index which is related to the time.

Let us first describe the system evolution under the effect of  a single operator (\ref{eq:heat_meas_block}). We suppose that the system and detector are initially described by the factorized density operator $\rho_0 = \rho_{S,0} \otimes \rho_{D}= \sum_{i_0, j_0,\lambda,\lambda'} \rho_{i_0 j_0}  \rho_{\lambda\lambda'} \ketbra{i_0,p}{j_0,\lambda'}$ where $ \rho_{i_0 j_0}$ and $\rho_{\lambda\lambda'}$ are the system and detector density matrix elements.
It is convenient to write the system density matrix in the eigenbasis of the initial Hamiltonian Hamiltonian $H_S^{0}$. In the following it will be useful to use write the evolution in the time-dependent basis  of the time-dependent Hamiltonian $H_S^{s}$ denoted with $\{ \ket{i_s}\}$, $\{ \ket{j_s}\}$ , $\{ \ket{n_s}\}$ and $\{ \ket{m_s}\}$.
Denoting with $U^0=\exp\{-i \Delta t H^{0}_S \}$, the evolution in Eq. (\ref{eq:heat_meas_block}) reads

The operator sequence of system evolution $\exp\{-i \Delta t H^{0}_S \}$, first system-detector coupling $\exp\{ i  \frac{\chi}{2} H_S^0 \otimes H_D\}$, dissipation and  second system-detector coupling $\exp\{- i  \frac{\chi}{2} H_S^0 \otimes H_D\}$ leads to the evolution 
\begin{eqnarray}
\rho_0  &\rightarrow& \sum_{} \rho_{\lambda\lambda'} U^0_{i_0 i_0} \rho_{i_0 j_0} (U^0)^\dagger_{j_0 j_0}    \ketbra{i_0,\lambda}{j_0,\lambda'} \nonumber \\
&\rightarrow& \sum_{} \rho_{\lambda\lambda'} e^{ i  \frac{\chi}{2} (\lambda \epsilon_{i_0}- \lambda' \epsilon_{j_0} )}  U^0_{i_0 i_0} \rho_{i_0 j_0} (U^0)^\dagger_{j_0 j_0}    \ketbra{i_0,\lambda}{j_0,\lambda'} \nonumber \\
&\rightarrow& 
\sum_{} \rho_{\lambda\lambda'} e^{ i  \frac{\chi}{2} (\lambda \epsilon_{i_0}- \lambda' \epsilon_{j_0} )}  
M^{k,0}_{n_0i_0}  U^0_{i_0 i_0} \rho_{i_0 j_0} (U^0)^\dagger_{j_0 j_0}(M^{k,0})^\dagger_{j_0m_0} \times \nonumber \\
&&
\ketbra{n_0,\lambda}{m_0,\lambda'} \nonumber \\
&\rightarrow&
\sum_{} \rho_{\lambda\lambda'} e^{ i  \frac{\chi}{2} [\lambda (\epsilon_{i_0} - \epsilon_{n_0})- \lambda' (\epsilon_{j_0}- \epsilon_{m_0}) ]} \times \nonumber \\
&& M^{k,0}_{n_0i_0}  U^0_{i_0 i_0} \rho_{i_0 j_0} (U^0)^\dagger_{j_0 j_0}(M^{k,0})^\dagger_{j_0m_0} \ketbra{n_0,\lambda}{m_0,\lambda'}
\label{app_eq:dissipative_first_step}
\end{eqnarray}
where at every step the sum must be considered on every index in the density matrix element.
Notice that, because we are decomposed the initial density operator in the eigenbasis of $H_S^0$, the first dynamical operator contributed only to phase factors, e.g., $U^0_{i_0 i_0} = \exp\{-i \Delta t  \epsilon_{i_0}\}$.

Generalizing this result to the dynamics discretized in $N$ steps, denoting with $n_{s-1} = l$ and $m_{s-1} = r$ and using the short notation $\mathcal{A}^{P_1}_{n_N i_0} = \Big( \Pi_{s=0}^N  M^{k,s}_{n_s i_s} U^l_{i_s n_{s-1}}\Big)$ and $(\mathcal{A}^{P_2}_{j_0 m_N})^\dagger = \Big( \Pi_{s=0}^N (U^l)^\dagger_{m_{s-1} j_s} (M^{k,s})^\dagger_{j_s m_s} \Big)$, we have
\begin{eqnarray}
\rho_N
 &=& \sum_{P_1,P_2} \rho_{\lambda\lambda'}  
e^{i  \frac{\chi}{2} [\lambda q_{P_1} - \lambda' q_{P_2} ]} \mathcal{A}^{P_1}_{n_N i_0} \rho_{i_0 j_0}
(\mathcal{A}^{P_2}_{j_0 m_N})^\dagger \times \nonumber \\ 
&& \ketbra{n_N,p}{m_N,\lambda'}
\label{app_eq:rhoN_heat}
\end{eqnarray}
where $\mathcal{A}^{P_1}_{n_N i_0}$ is the probability amplitude of path $P_1:l \rightarrow i_0 \rightarrow n_0 \rightarrow i_1 \rightarrow .... \rightarrow n_N$ and $\mathcal{A}^{P_2}_{j_0 m_N}$ is the probability amplitude for the path $P_2:r \rightarrow j_0 \rightarrow m_0 \rightarrow j_1 \rightarrow .... \rightarrow m_N$.
They are related to dissipated heat $q_{P_1} =\sum_{s }(\epsilon_{i_s} - \epsilon_{n_s})$ and $q_{P_2} = \sum_s (\epsilon_{j_s}- \epsilon_{m_s})$, respectively.

From the dynamics of the density operator, we can determine the \CharFun and \ProbFun for the dissipated heat.
The \ProbFun for the variation of internal energy and work can be determined by the above one as discussed in the main text.

It is worth making an additional comment.
The \ProbFun for the internal energy variation reads (see main text)
\begin{equation}
 \Prob(\Delta U) = \sum_{P_1,P_2}  
 \mathcal{A}^{P_1}_{n_N i_0} \rho_{i_0 j_0}
(\mathcal{A}^{P_2}_{j_0 n_N})^\dagger \delta \Big[\Delta U -\lambda \left(\epsilon_{n_N} - \frac{\epsilon_{i_0}+ \epsilon_{j_0}}{2} \right)\Big].
\end{equation}
As we can see, the energy exchanges depends only on $\epsilon_{i_0}$ and $\epsilon_{j_0}$

\section{Strong dissipation regime}
\label{app:strong_dissipation_dynamics}

We discuss the limit of strong dissipation for a two-level system.
In this case, the operator elements inthe operator sum-representation \cite{nielsen-chuang_book, Garcia-Perez2020} in the $\{ \ket{g_k}, \ket{e_k} \}$ basis are
\begin{eqnarray}
M^{0,s} &=& 
\left(
\begin{array}{cc}
 1 & 0 \\
 0 &  \sqrt{1-p} \\
\end{array}
\right) \nonumber \\
M^{1,s} &=&
\left(
\begin{array}{cc}
 0  & \sqrt{p} \\
 0& 0 \\
 \end{array}
\right).
\label{eq_app:operator_elements}
\end{eqnarray}
where the first  index is the operator one while $s$ represents the time index, i.e., the operator at time $t_s$.
At any time $t_s$, we have that $\rho_S(t_{s+1}) = M^{0,s} \rho_S(t_s) M^{0,s} + M^{1,s} \rho_S(t_s) M^{1,s}$.

In the case of strong dissipation, i.e., $p=1$, the system always completely relaxes to the instantaneous ground state of the Hamiltonian.
From Eq. (\ref{eq_app:operator_elements}), the dissipative operators elements can be written as $M^{0,s}_{ij} = \delta_{i g_s} \delta_{j g_s}$, $M^{1,s}_{ij} = \delta_{i g_s} \delta_{j e_s}$ and $(M^{1,s})^\dagger_{ij} = \delta_{i e_s} \delta_{j g_s}$.

First, we calculate $\Prob(\Delta U)$ in this limiting case. Since the system-detector couplings are performed only at the beginning and at the end of the evolution the discussion is greatly simplified.
The evolution can be obtained by Eq. (\ref{app_eq:dissipative_first_step}) recalling the after the dissipation there is no coupling with the operator $\exp\{- i  \frac{\chi}{2} H_S^0 \otimes H_D\}$.
We have 
\begin{eqnarray}
&& \sum_{} \rho_{\lambda\lambda'}  \rho_{i_0 j_0}   \ketbra{i_0,\lambda}{j_0,\lambda'}
\rightarrow \nonumber \\
&& \sum_{}\rho_{\lambda\lambda'} e^{ i  \frac{\chi}{2}  (\lambda\epsilon_{i_0} - \lambda'\epsilon_{j_0})} 
M^{k,0}_{n_0i_0}  \rho_{i_0 j_0} (M^{k,0})^\dagger_{j_0m_0} \ketbra{n_0,\lambda}{m_0,\lambda'}.
\label{app_eq:density_operator_second}
\end{eqnarray}
(Notice that in the last sum the dissipative operators are at time $t=0$ and summed on the $k$ index).

With a closer analysis, we have that $\sum_{k} M^{0,s}_{n_0i_0}  \rho_{i_0 j_0} (M^{0,s})^\dagger_{j_0m_0} =
  M^{0,0}_{n_0i_0}  \rho_{i_0 j_0} (M^{0,0})^\dagger_{j_0m_0}  + M^{1,0}_{n_0i_0}  \rho_{i_0 j_0} (M^{1,0})^\dagger_{j_0m_0} $.
By using the above explicit expression of the dissipative operators, in the first term all the indices are fixed to $g_0$, i.e., $i_0=j_0=n_0=m_0=g_0$, while in the second one we have $i_0=j_0=e_0$ and $n_0=m_0=g_0$.
The full sum in Eq. (\ref{app_eq:density_operator_second}) reads
\begin{equation}
 \sum \rho_{\lambda\lambda'}  \rho_{i_0 j_0}   \ketbra{i_0,\lambda}{j_0,\lambda'}
\rightarrow  \sum \rho_{\lambda \lambda'} \bar{\rho}_{g_0 g_0} \ketbra{g_0,\lambda}{g_0,\lambda'}.
 \label{app_eq:rho_first_dissipation}
\end{equation}
where 
\begin{equation}
 \bar{\rho}_{g_0 g_0} = \rho_{g_0 g_0} e^{ i  \frac{\chi}{2}  (\lambda - \lambda')\epsilon_{g_0}} +  \rho_{e_0 e_0}  e^{i   \frac{\chi}{2}  (\lambda- \lambda') \epsilon_{e_0} }.
 \label{app_eq:bar_rho_00}
\end{equation}

As expected, the system is in the ground state after the dissipation.
The first line is just an alternative writing to point out the origin of the terms that sum up to $ \bar{\rho}_{g_0 g_0}$.

Since we are now interested in the internal energy, only the final state of the system is important, that is when the second system-detector coupling takes place.
The intermediate dynamics is complex but we know that, because of the strong relaxation, the system eventually ends up in the $\ket{g_N}$ state.
In other terms, the system state $\ket{g_0}$ undergoes a sequence of excitations (due to the drive) and relaxations (due to the environment).
However, it eventually relaxes to the final ground state $\ket{g_N}$ so that the effective dynamics is simply $\ket{g_0} \rightarrow \ket{g_N}$.

With these observations, we can immediately write the evolution of (\ref{app_eq:rho_first_dissipation}) that becomes $\rho_N = \sum_{\lambda \lambda'} \rho_{\lambda \lambda'} \bar{\rho}_{g_0 g_0} \ketbra{g_N ,\lambda}{g_N,\lambda'}$ and, after the second system-detector coupling, we obtain
\begin{equation}
  \rho_N = \sum_{\lambda \lambda'} \rho_{\lambda \lambda'} \bar{\rho}_{g_0 g_0} e^{-i   \frac{\chi}{2}  (\lambda- \lambda') \epsilon_{g_N} } \ketbra{g_N ,\lambda}{g_N,\lambda'} 
\end{equation}

Assuming that $\epsilon_{g_N} = \epsilon_{g_0}$ and using Eq. (\ref{app_eq:bar_rho_00}), the corresponding probability density function reads (recalling that we have to take $\lambda'=-\lambda=-1$ and then calculate the Fourier transform)
\begin{equation}
  \Prob(\Delta U) = \rho_{g_0 g_0} \delta(\Delta U) +  \rho_{e_0 e_0} \delta \left[ \Delta U - (\epsilon_{e_0}-\epsilon_{g_N}) \right].
\end{equation}
This has an immediate interpretation.
If the system starts and ends in the ground state there is no variation of internal energy. This occurs with probability $\rho_{g_0 g_0}$.
If the system starts in the excited state and ends in the ground state the energy variation is $\epsilon_{e_0}-\epsilon_{g_N}$ and this process occurs with probability $\rho_{e_0 e_0}$.

More importantly, the half-energy exchanges are not possible and the allowed exchanges are associated to positive probabilities ($\rho_{g_0 g_0}$ and $\rho_{e_0 e_0}$).
Since these were the signature of a quantum process and were associated with the violation of the Leggett-Garg inequality, we deduce the presence of the environment has made the evolution classical.

With a similar approach, it is possible to calculate the heat distribution $\Prob(Q)$ in the strong dissipative limit.
The general first evolution and dissipation are given by Eq. (\ref{app_eq:dissipative_first_step}).
Following the calculation for  $\Prob(\Delta U)$, we impose the constraints on the $M$ operators that impose $i_0=j_0=n_0=m_0=g_0$ for the first term and $i_0=j_0=e_0$, $n_0=m_0=g_0$ for the second term.
In this way, we obtain the ground state matrix element ($\ketbra{g_0,\lambda}{g_0,\lambda'}$) after the dissipation
\begin{eqnarray}
&& \sum \rho_{\lambda\lambda'}  \rho_{i_0 j_0}   \ketbra{i_0,\lambda}{j_0,\lambda'}
\rightarrow \nonumber \\
&& \sum_{\lambda\lambda'} \rho_{\lambda\lambda'} \left( \rho_{g_0 g_0} + M^{1,0}_{g_0 e_0} \rho_{e_0 e_0} (M^{1,0})^\dagger_{e_0 g_0}  e^{i \frac{\chi}{2} (\lambda- \lambda') (\epsilon_{e_0} - \epsilon_{g_0})} \right) \times \nonumber \\
&& ~~~\ketbra{g_0,\lambda}{g_0,\lambda'}
 \label{app_eq:rho_heat_1}
\end{eqnarray}
The first term corresponds to the absence of dissipation since it is associated with the transformation is $g_0 \rightarrow g_0$.
The second is related to the transition $e_0 \rightarrow g_0$ (due to the $M^1$ operators) and associated to the energy exchange $\epsilon_{e_0} - \epsilon_{g_0}$ that appears in the exponent.

The following steps are analogous.
After the application of the $U^1$ operator, the new density operator in the $\{\ket{i_1} \}$ basis reads $\sum_{i_1,j_1} \rho_{\lambda\lambda'}  \rho_{i_1 j_1}   \ketbra{i_1,\lambda}{j_1,\lambda'}$ with $ \rho_{i_1 j_1} = U^1_{i_1 i_0}\rho_{i_0 j_0} (U^1)^\dagger_{j_0 j_1}$.
Analogously to the above calculation [Eq. (\ref{app_eq:bar_rho_00})], the matrix element of the system density operator after the second dissipation is
\begin{equation}
\bar{\rho}_{g_1 g_1} =  \rho_{g_1 g_1} + M^{1,1}_{g_1 e_1} \rho_{e_1 e_1} (M^{1,1})^\dagger_{e_1 g_1} e^{i  \frac{\chi}{2}(\lambda- \lambda') (\epsilon_{e_1} - \epsilon_{g_1})}.
\end{equation}

It is interesting to explicitly write this using the above results. We have four contributions
\begin{eqnarray}
&& U^1_{g_1 g_0}\rho_{g_0 g_0} (U^1)^\dagger_{g_0 g_1} \nonumber \\
&+&  U^1_{g_1 g_0} M^{1,0}_{g_0 e_0} \rho_{e_0 e_0} (M^{1,0})^\dagger_{e_0 g_0} (U^1)^\dagger_{g_0 g_1} e^{i  \frac{\chi}{2} (\lambda- \lambda') (\epsilon_{e_0} - \epsilon_{g_0})}
 \nonumber \\
 &+& 
 M^{1,1}_{g_1 e_1} U^1_{e_1 g_0} \rho_{g_0 g_0}  (U^1)^\dagger_{g_0 e_1} (M^{1,1})^\dagger_{e_1 g_1} e^{i  \frac{\chi}{2} (\lambda- \lambda') (\epsilon_{e_1} - \epsilon_{g_1})} 
 \nonumber \\
 &+& 
M^{1,1}_{g_1 e_1} U^1_{e_1 e_0} M^{1,0}_{g_0 e_0} \rho_{e_0 e_0} (M^{1,0})^\dagger_{e_0 g_0}  (U^1)^\dagger_{e_0 e_1} (M^{1,1})^\dagger_{e_1 g_1} \times \nonumber \\
&& e^{i  \frac{\chi}{2} (\lambda- \lambda') (\epsilon_{e_1} - \epsilon_{g_1} + \epsilon_{e_0} - \epsilon_{g_0})}.
\end{eqnarray}
The first term corresponds to the transition $g_0 \rightarrow g_1$ and no dissipated heat.
The second one corresponds to the transition $e_0 \rightarrow g_0 \rightarrow g_1$ and dissipated heat $\epsilon_{e_0} - \epsilon_{g_0}$  during the first dissipative process.
The third one corresponds to the transition $g_0 \rightarrow e_1 \rightarrow g_1$ and dissipated heat $\epsilon_{e_1} - \epsilon_{g_1}$  during the second dissipative process.
The fourth one corresponds to the transition $e_0 \rightarrow g_0 \rightarrow e_1 \rightarrow g_1$ and dissipated heat $\epsilon_{e_0} - \epsilon_{g_0}$ and $\epsilon_{e_1} - \epsilon_{g_1}$  during both the dissipative processes.

By extension, the full dissipative evolution is the sum over all the dissipative paths. We obtain 
\begin{eqnarray}
\rho_N &=&  \sum_{P_1} \rho_{\lambda\lambda'}  
e^{i  \chi (\lambda-\lambda') q_{P_1}} \mathcal{A}^{P_1}_{g_N i_0} \rho_{i_0 i_0}
(\mathcal{A}^{P_1}_{i_0 g_N})^\dagger  \times \nonumber \\
 && ~~\ketbra{g_N,p}{g_N,\lambda'} 
\label{app_eq:rhoN_heat_strong_dissipation}
\end{eqnarray}
where $P_1:i_0 \rightarrow n_0 \rightarrow i_1 \rightarrow .... \rightarrow n_N$ and $i_k, n_k=g_k, e_k$.

There are important differences between the result for any dissipation strength in Eq. (\ref{eq:rhoN_heat}) and this one.
First, in Eq. (\ref{app_eq:rhoN_heat_strong_dissipation}) only the diagonal terms of the density operator appear $\rho_{i_0 i_0}$. This implies that the effect of the initial coherences are destroyed by the presence of the environment.
Second and more important, there is only a path $P_1$ in the path summation.
The second path $P_1$ which should cause the quantum interference in the evolution disappears because of the environmental effect.
This means that the term $\mathcal{A}^{P_1}_{g_N i_0} \rho_{i_0 i_0} (\mathcal{A}^{P_1}_{i_0 g_N})^\dagger$ can be interpreted as the probability to go from $i_0$ to $g_N$.
This is not possible with different paths $P_1$ and $P_2$.

The corresponding probability density distribution is 
\begin{equation}
  \Prob(Q) =  \sum_{P_1} \mathcal{A}^{P_1}_{g_N i_0} \rho_{i_0 i_0}
(\mathcal{A}^{P_1}_{i_0 g_N})^\dagger  \delta \left[ Q - q_{P_1} \right]
\label{app_eq:PQ}
\end{equation}
Since $\mathcal{A}^{P_1}_{g_N i_0} \rho_{i_0 i_0} (\mathcal{A}^{P_1}_{i_0 g_N})^\dagger$ are now probabilities, this is a positive probability density function.

As discussed in the main text, the protocol to determine the \ProbFun of the work is similar to the one used in the heat apart from two additional system-detector coupling at the beginning and the end of the evolution.
Therefore, the $\Prob(Q)$ can be immediately obtained from Eq. (\ref{app_eq:PQ})
\begin{equation}
  \Prob(W) =  \sum_{P_1} \mathcal{A}^{P_1}_{g_N i_0} \rho_{i_0 i_0}
(\mathcal{A}^{P_1}_{i_0 g_N})^\dagger  \delta \left[ Q - (q_{P_1} - \epsilon_{e_0}+\epsilon_{g_N}) \right]
\label{app_eq:PQ}
\end{equation}


%



\end{document}